\documentclass[a4paper,fleqn]{cas-dc}

\usepackage[authoryear]{natbib}

\usepackage{defs}
\begin{document}
\let\WriteBookmarks\relax
\def\floatpagepagefraction{1}
\def\textpagefraction{.001}

\shorttitle{SHARP view on protoclusters}    

\shortauthors{M. Polletta et al.}  

\title [mode = title]{Fueling and feedback mechanisms at the nodes of the cosmic web}  




\author[1]{Maria del Carmen Polletta}[orcid=0000-0001-7411-5386]
\cormark[1]
\ead{maria.polletta@inaf.it}
\credit{Writing - original draft, Conceptualization, Methodology}
\affiliation[1]{organization={INAF – Istituto di Astrofisica Spaziale e Fisica Cosmica Milano},
            addressline={Via A. Corti 12}, 
            city={Milan},
            postcode={20133}, 
            state={MI},
            country={Italy}}

\author[2,3]{Gabriella De Lucia}
%[orcid=0000-0002-6220-9104]
%\ead{gabriella.delucia@inaf.it}
\credit{Writing - review}
\affiliation[2]{organization={INAF - Astronomical Observatory of Trieste},
            addressline={Via G. B. Tiepolo 11}, 
            city={Trieste},
            postcode={34143}, 
            state={TS},
            country={Italy}}
\affiliation[3]{organization={Institute for Fundamental Physics of the Universe},
            addressline={Via Beirut 2}, 
            city={Trieste},
            postcode={34151}, 
            state={TS},
            country={Italy}}

\author[4]{Anna R. Gallazzi}
%[orcid=0000-0002-9656-1800]
\affiliation[4]{organization={INAF-Osservatorio Astrofisico di Arcetri},
            addressline={Largo Enrico Fermi 5},
            city={Firenze},
            postcode={50126},
            state={FI},
            country={Italy}}
%\ead{anna.gallazzi@inaf.it}
\credit{Writing - review}

\author[1]{Chiara Mancini}
%[orcid=0000-0002-4297-0561]
%\ead{chiara.mancini@inaf.it}
\credit{Writing - review}

\cortext[1]{Corresponding author:}

\begin{abstract}
The environment plays a key role in shaping how galaxies form and evolve. Galaxies in dense nodes of the cosmic web are thought to grow and quench earlier, and faster and become more massive than those in the field.  To understand the physical drivers of this environmental effect, we must probe the most crowded regions of the Universe at the epoch when growth was at its peak and the transition to quiescence was triggered, around 10 billion years ago ($z\sim2$).  This period saw the downturn of the cosmic star-formation and black-hole accretion histories, the quenching and morphological transformation of massive galaxies, and the virialisation of the first clusters.  Several processes might be at play: stellar and AGN feedback, reduced gas accretion, disk instabilities, morphological quenching, interactions, and ram-pressure stripping.
The ELT/SHARP instrument, with its sensitivity, spectral resolution, wavelength coverage, and multiplexing capabilities over a wide field, is ideally suited to study these mechanisms by targeting multiple members of dense structures simultaneously.  Cluster and protocluster cores at $z\sim2$ span roughly 1\arcmin\ and host about ten massive (M$_{\rm star}>10^{10.5}$\,\msun) galaxies.  VESPER can deliver spatially resolved gas and stellar kinematics, map recent and past star formation, identify companions, inflows, outflows, shocks, and AGN activity for the most massive core members. With 80\,hr of VESPER time, we can obtain this type of data for about 60 galaxies selected from the densest regions of five clusters at $1.5<z<1.7$ and five protoclusters at $2<z<2.5$ spanning the evolutionary phases of maximal growth and rapid decline. Such a sample would permit to trace the evolution from protoclusters to virialised clusters and identify the environmental processes responsible for their rapid transformations.
\end{abstract}

% Keywords

\maketitle

% Main text
\section{Scientific Rationale}\label{sec:intro}
\begin{figure}%[ht]
    \centering
    \includegraphics[width=0.5\textwidth]{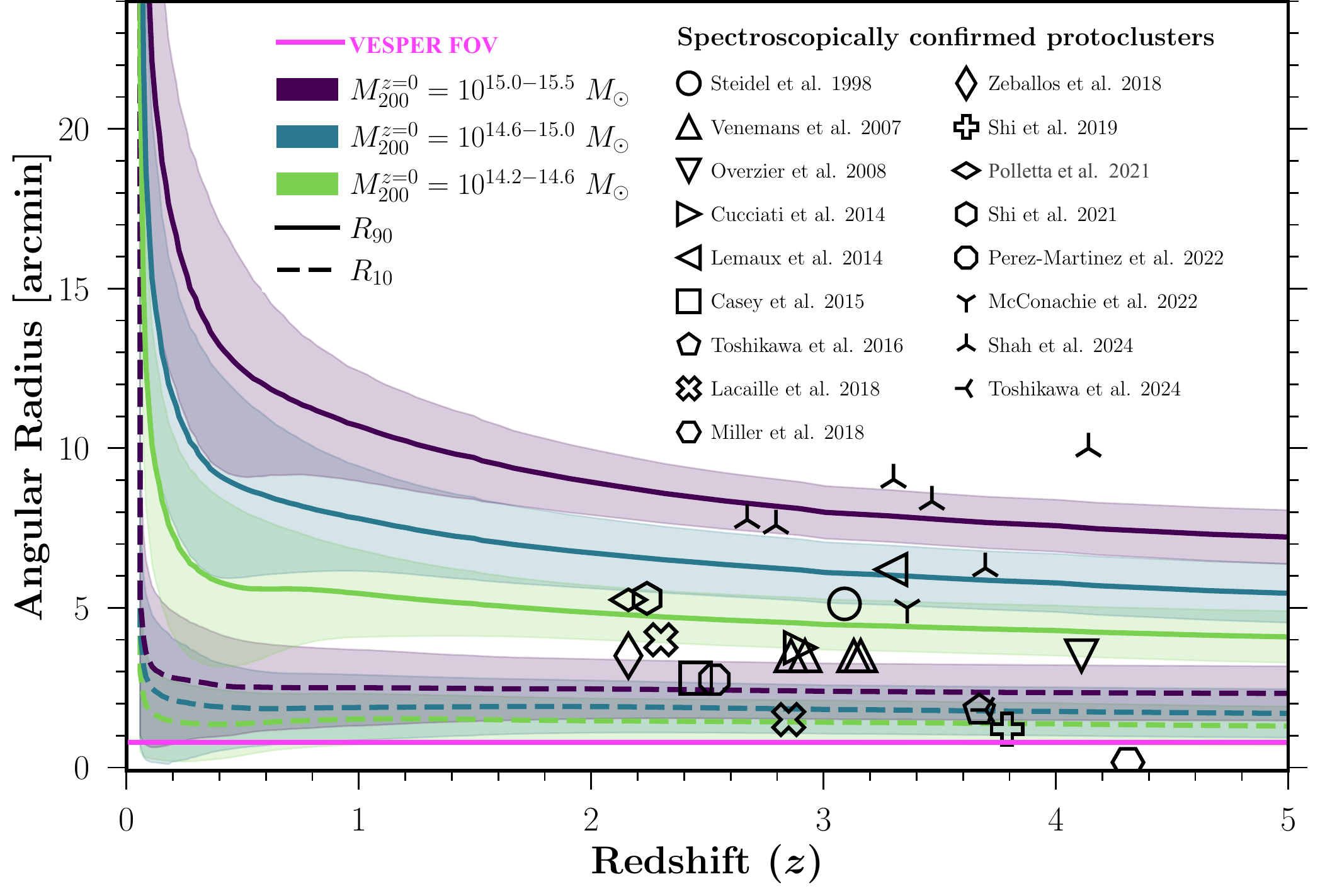}
   \caption{Protocluster size evolution in units of arcmin, characterized by R$_{\rm 90}$ (solid lines) and R$_{\rm 10}$ (dashed lines), which represent the radii, measured relative to the center of mass of the ensemble of protocluster members at the observed redshift, that enclose 90\% and 10\%, respectively, of the total stellar mass of the  protocluster galaxies with M$_{\rm star}>10^{8.5}$\,\msun\ that will reside within R$_{\rm 200}$ at $z=0$. The purple, blue, and green lines indicate the mean R$_{\rm 90}$ (solid) and R$_{\rm 10}$ (dashed) size values for protoclusters in three bins of $z=0$ halo mass (purple: $M_{\rm 200}=10^{15.0-15.5}$\,\msun, blue:  $M_{\rm 200}=10^{14.6-15.0}$\,\msun, and green: $M_{\rm 200}=10^{14.0-14.6}$\,\msun), with bands showing the corresponding 68\% quantiles. Open markers show the angular sizes of spectroscopically confirmed protoclusters from the literature with $>$10 members at $2<z<5$. The magenta line represents the FOV of VESPER (i.e., 47\arcsec\ on average). Figure adapted from \citet{baxter25}. }
    \label{fig:pc_size_vs_z}
\end{figure}

Galaxies are not randomly distributed on the sky, but they gather in a ``cosmic web'', an intricate ensemble of filaments, nodes, and voids. Only recently the importance of the cosmic web has been recognized and numerous studies have tried to determine how the location of a galaxy within it affects its evolution \citep[see e.g.,][]{gao26}. High-redshift galaxies in overdense environments experience an earlier and more rapid growth \citep{steidel05,hatch11,koyama13,sun25}, and decline in star formation \citep{kubo21,shi21,alberts22a,mcconachie22} than those in the field.  This leads to a dominant population of massive and passive galaxies in cluster cores at low redshifts \citep{dressler80}.  The environmental conditions that favored their growth are thought to be larger reservoirs of gas and a higher probability of galaxy encounters. However, the physical mechanisms that regulate this accelerated evolution, and the conditions that enable their occurrence are not fully understood. To address these questions, many observational studies have been carried out with the goal of finding the precursors of galaxy clusters, the so-called galaxy protoclusters,  and of investigating the properties of their inhabitants. A significant advancement has been made in the past decades with tens of protoclusters discovered up to high redshifts \citep[e.g.,][]{li25}, but our understanding remains largely phenomenological, with investigations primarily limited to characterizing differences between protoclusters and coeval field populations, while the physics driving this differential evolution remains poorly understood.\\
Galaxy protoclusters are often spotted as galaxy overdensities around high redshift quasi stellar objects \citep[QSOs;][]{mignoli20}, radio galaxies \citep[RGs;][]{galametz13} or submillimeter galaxies \citep[SMGs);][]{miller18,oteo18,polletta21}, or thanks to highly sampled spectroscopic observations \citep{cucciati18,lemaux14b}. The galaxy members are usually identified through narrow band imaging observations as \lya\ or \Ha\ emitters \citep[LAE and HAE;][]{shi2019,koyama13,koyama21}, or through spectroscopic observations in the visible, in the near-infrared (NIR) or in the millimeter \citep[mm; e.g.,][]{capak11,ivison13,cucciati14,lemaux18,kneissl19,polletta22,hill25,laporte22}. 
This heterogeneous sampling results in incomplete and biased membership determinations, and in differences in characterizing the members' general properties (such as stellar mass, star formation rate, gas fraction, AGN contribution), as well as in deriving those of the protoclusters as a whole, such as their current and expected $z=0$ total halo mass. Since a cluster assembly history can be quite diverse, with some clusters accumulating their mass rapidly at early times, while others growing more gradually, an estimate of their final halo mass based on the observed properties, such as the protocluster total stellar mass, star formation rate (SFR) or richness, can be highly uncertain \citep{gouin22,remus23}. The most common way to estimate the final ($z=0$) mass of a protocluster it is to identify a similar structure in a large volume cosmological simulation and follow its evolution up to $z=0$ \citep[see e.g.,][]{gouin22,ata22}. This method obviously confides on the reliability of the simulation in reproducing the observations. \\
Large volume cosmological simulations have made noticeable progress in characterizing these large scale structures and have guided the observers in identifying and characterizing them \citep{pillepich18,nelson19,springel05a,boylan09,euclid_pc,forever22_pc_sims}. According to simulations, the progenitors of massive clusters extend across several megaparsecs \citep[up to $12-15$\,cMpc, or $\sim8-10$\arcmin\ at $z\sim2$ for the most massive ones ($M^{z=0}>10^{15}$\,\msun);][]{muldrew15,shimakawa18,chiang17}.  Their size is expected to scale with the mass of the descendant clusters \citep[e.g.,
from $\sim$4\arcmin\ to $\sim$10\arcmin\ for haloes with masses from $\sim10^{14.2}$\,\msun, to $\sim10^{15.5}$\,\msun\ at $z\sim2$;][]{kikuta19,baxter25}. Observationally, protoclusters span several arcmin on the sky
\citep{perez_martinez23,shi21,polletta21,kneissl19,hill25}. Overall, there is a fair agreement between the range of observed and predicted angular extents of protoclusters with different final masses from simulations \citep[e.g.,][]{tngcluster} and the measured size of several confirmed protoclusters. However, many of the existing spectroscopically confirmed protoclusters do not fully capture the expected size occupied by their progenitors. This mismatch is illustrated in Fig.~\ref{fig:pc_size_vs_z} \citep{baxter25}, where we compare the extent of known protoclusters with the  size of simulated structures enclosing 90\% of stellar mass. \\
A major step forward in protocluster studies will be made possible by the acquisition of spectroscopic data over large areas such as those of the ever-growing spectroscopic catalogs in legacy fields \citep{khostovan26}, and large spectroscopic surveys, such as DESI \citep{desi}, or those that are being or will be carried out by the James Webb Space Telescope \citep[JWST;][]{morishita23,li25}, Euclid \citep{euclid_pc,bohringer25_euclid}, LSST \citep{lsst,gully24}, and Nancy Grace Roman
Space Telescope \citep[Roman;][]{rudnick23}. These will pinpoint overdensities, map their full extent, and enable population level analyses of at least some member types. However, to fully understand the environmental-related physical processes that galaxies experience in overdense environments, a complementary approach is also needed, that is zooming into the densest regions and studying in detail the physical processes that govern the rapid evolution of the member galaxies.
\subsection{The growth and decline of galaxies in overdense regions}

The cores of protoclusters at $z\gtrsim2$, besides being denser, often host members that are, on average, more massive, and star-forming and with a higher AGN fraction than the full protocluster population, suggesting an environmental effect that favors galaxy growth and AGN activity \citep{koyama21,polletta21,oteo18,miller18,digby10,lehmer13,polletta21,Tozzi2022,vito24}. At $z\lesssim$2, a transition, from this active growing phase to a passively evolving phase is typically observed. Indeed the quiescent fraction in clusters rapidly increases from $z\sim2$ to later times, although significant variations are observed across clusters, implying intrinsic differences or possibly quenching timescales that are shorter than those captured by the observations \citep{newman14,strazzullo18,papovich18}. \\
This transition corresponds to the epoch when the cosmic star formation and BH accretion histories start declining after reaching their highest values \citep{madau14,kim24}, and when galaxy structures start collapsing and virialising. This suggests a link between the environment and the mechanisms that regulate the activity level within massive galaxies.
The enhanced activity level at $z\gtrsim2$ is usually attributed to inflows of gas from the cosmic web \citep{umehata19} and to an elevated likelihood of mergers in dynamically young regions \citep{alberts22a}. Both processes can fuel the star formation activity and feed the super massive black hole (SMBH) at the center of galaxies, lightening up the AGN. 
Similarly, a decrease in efficiency of gas accretion and of mergers might be at the origin of the decline in star formation at $z<2$ and explain the dominant fraction of massive quiescent galaxies observed in galaxy clusters at $z<1.5$.  According to simulations, the cold gas from the cosmic web can penetrate even massive haloes (i.e., $\sim$10$^{11.8}$\,\msun) and reach the central regions through narrow streams at $z\gtrsim2$, but, at $z\lesssim2$, the inflowing gas is shock-heated and becomes unavailable to fuel star formation \citep{dekel06,dekel09}. Also merger events become less frequent at $z\lesssim2$ because velocities increase when the core virializes and the ICM temperature rises \citep{liu25}. Other mechanisms can also intervene in massive halos \citep[e.g., gas stripping, overconsumption, disruption;][]{alberts24} and cause a decline in star formation.\\
This transitional phase at $z\sim2$ and the role of the different processes that affect the gas capability to form stars and feed the AGN are still poorly constrained by the observations and, consequently, difficult to simulate. For example, most simulations fail in reproducing the extremely active members (i.e., those with SFR$>$100s\,\msun\,yr$^{-1}$; \citealt{bassini20,lim21,gouin22}, but see the TNG-Cluster simulation; \citealt{tngcluster}), that are instead found in many protoclusters \citep{casey15,umehata15,oteo18,miller18,polletta22}. Also the mass, kinematics and rates of the inflowing and outflowing gas predicted by simulations suffer large variations, especially at small galactocentric distances ($<$200\,kpc), even if they are run with the same initial conditions \citep[see e.g., CAMELS;][]{medlock25,crain23,wright24}. There are also large differences in the number density of quiescent galaxies as a function of redshift produced by current hydrodynamical simulations and semi-analytical models \citep{lagos25}.\\
To make progress on our understanding of the environment-related mechanisms responsible for the growth and the decline of the star formation activity, we need to acquire observables that trace the gas properties and the activity level in massive galaxies in overdense regions spanning the evolutionary phases of maximal growth and rapid decline. \\
The SHARP instrument \citep{sharp24} designed for the Extremely Large Telescope (ELT) offers the spectral coverage and resolution, sensitivity and spatial resolution that will enable to observe the core members of protoclusters and collapsing clusters and study the mechanisms that regulate their star formation activity and cause their quenching.
\section{The SHARP program on galaxy protoclusters}

We propose an observational program to carry out with SHARP to study the population of massive ($>10^{11}$\,\msun) galaxies living in the most crowded regions of the Universe at $z\sim2$. 

\subsection{Scientific goal}

The main goal of this program is to determine the environmental processes driving growth and quenching in massive ($>10^{11}$\,\msun) cluster and protocluster members. 
We will address these questions by studying a large sample of galaxies located in extremely dense environments. In particular, we plan to observe with VESPER the core regions of clusters at $1.5<z<1.7$, and of protoclusters at $2.0<z<2.5$. VESPER will provide resolved maps of key spectral features that will probe the distribution, the chemistry and the kinematics of the gaseous and stellar components of these massive galaxies.\\
We will assess the role of gas accretion and of mergers in fueling the star formation activity in the star-forming members by comparing the presence of inflowing gas and of morphological disturbances with the distribution of specific SFR, and stellar ages. We will examine the distribution of stellar ages and SFR density in the transitioning members to determine if quenching is "outside-in" (stripping, or strangulation) or "inside-out" (AGN feedback). In the presence of an outflow, we will compare its mass rate with the SFR and the AGN power to assess its origin and impact. We will compute several parameters such as the dynamical status, the energetics from stellar radiation, AGN activity and shocks, the morphology (bulge-to-total ratio, level of disturbance, star-forming clumps contribution), the metallicity and its distribution, the presence of inflows, outflows, and the dust properties.
The mass cut and environment selection of the two sub-samples (cluster and protocluster members) will allow us to analyze statistical trends among the various parameters neglecting the effect of redshift, mass and environment. By cross-correlating them, we will identify the environmental and internal mechanisms that modulate the activity level of massive galaxies in the two selected sub-samples. The comparison between the two sub-samples will reveal the mechanisms that affect their evolution.

\subsection{Measurements and analysis plan} 
\label{sec:objectives}

\begin{figure}%[ht!]
    \centering
    \includegraphics[width=0.23\textwidth]{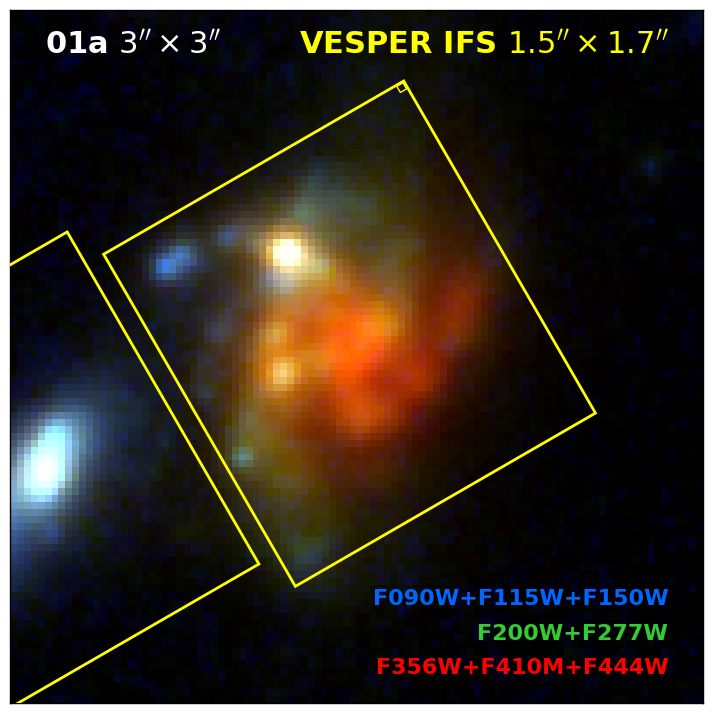}
    \includegraphics[width=0.23\textwidth]{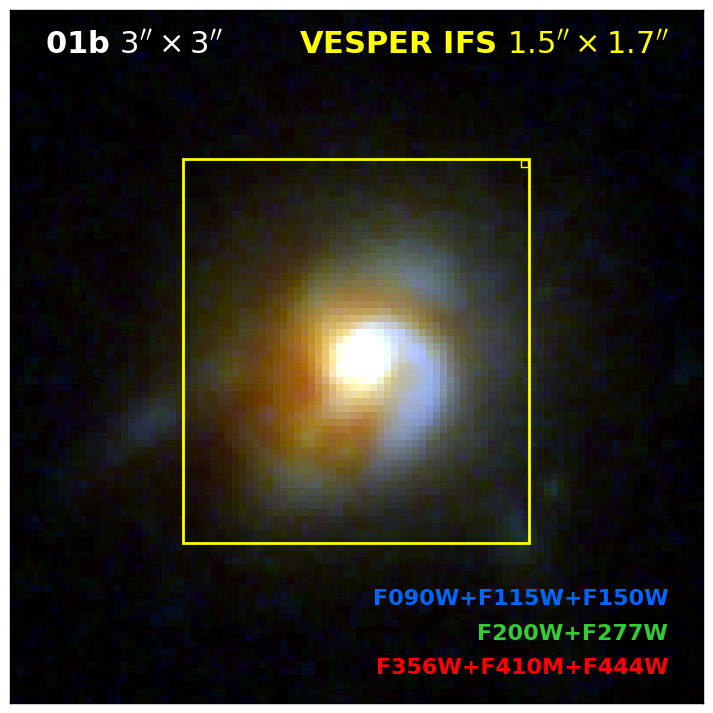}
   \caption{JWST multi-band 3\arcsec$\times$3\arcsec\ images of the starburst galaxies G191 01a and 01b at $z\sim2.5$
\citep{polletta24} and the FOV of a VESPER IFS (yellow rectangle).}
    \label{fig:jwst_images}
\end{figure}

With VESPER, we plan to obtain the following measurements and carry out the analysis described below for each member galaxy: (1) a morphological study, (2) a spectral analysis, (3) build resolved maps of stellar ages, SFR density, stellar mass and extinction, and (4) a kinematic investigation of the gaseous and stellar components.\\
\textbf{(1) Morphological analysis: }
We will model the light of each galaxy to look for signs of morphological transformation from disks to bulge dominated \citep[see e.g.,][]{golden_marx26}, and to identify different sub-components.  In Fig.~\ref{fig:jwst_images}, we show two examples of $z\sim2$ galaxies imaged with JWST located in an overdensity \citep{polletta24}. The multi-band images, created using Trilogy \citep{trilogy}, show color variations across each galaxy that highlight star forming clumps, dusty regions, spiral arms, and faint extended structures.  The complexity of these images illustrates the need of decomposing their integrated light in order to reveal the different components and their origin.  The VESPER Integral Field Selector (IFS) spaxel size (31\,mas, equivalent to $\lesssim300$\,pc at the selected redshifts) is ten times smaller than the expected average effective radius of star-forming galaxies \citep[i.e., $R_{\rm e}\simeq3$\,kpc;][]{vanderwel25} thus permitting to resolve the galaxy light in hundreds of spaxels and identify the different regions such as the nucleus, the bulge, the disk, star-forming clumps and other sub-structures within a single IFS. With this approach, we will also look for signs of merging activity and interactions that might drive the intense star formation activity observed in some members \citep{tran10,santos14,polletta24}.\\
\textbf{(2) Spectral analysis} Visible spectra are rich of features that can serve to determine the different sources of ionization (AGN radiation, star formation, shocks), and as diagnostics for several key parameters, such as the ionization parameter \citep[e.g., from the \oiiib/\hbeta\ ratio;][]{osterbrock06,liu13}, the electron density \citep[e.g., from the \sii\ doublet;][]{li25_ne}, the metallicity \citep{maiolino19}, stellar ages \citep[hydrogen or metals absorption features;][]{leitherer10}, neutral gas \citep[Na\,I\,D absorption feature;][]{davies24}, and to identify shocks \citep[from the \oii/\oiiib\ ratio and the \gion{FeV}{2} $\lambda$5158 line;][]{zakamska14}.
These measurements will permit to disentangle the contribution and the impact of shocks, young stars radiation, and nuclear radiation by applying the BPT diagnostic diagram \citep{bpt_diagram}.\\
Another important diagnostic is provided by the gas metallicity because it is highly dependent on environmental effects (e.g., galaxy encounters, the intergalactic medium pressure), and can inform on accretion of pristine or recycled gas \citep{shimakawa15,chartab21}. 
Few studies have investigated the metallicity in protocluster galaxies.  Some report higher gas metallicities than coeval field galaxies \citep{shimakawa15,perez_martinez23,wang25_MQN01,li25}, others lower values \citep{sattari21,chartab21,polletta21,zhou25}, and some find a mass-dependent effect, with a metallicity deficit at high masses and an enhancement at low masses \citep{arribas24,yang26}. 
The discordant results obtained so far might be due to the hetereogeneous selection of protoclusters and of their members, to the use of integrated values, or to short timescale variations. Resolved metallicity maps of a sufficiently large sample are needed to determine the range of metallicities in protocluster massive galaxies, measure gradients and reveal the mechanisms that regulate it. \\
\textbf{(3) Resolved SED fitting: }
Resolved spectro-photometric fitting will provide resolved maps of stellar mass, ages, SFRs, and dust extinction \citep[see e.g.,][]{kamieneski24,polletta24}. These maps will be crucial to identify the regions where star formation activity is enhanced and those where it is quenched and determine the cause. SHARP wavelength coverage permits to study the effects of dust extinction and the dust properties. In particular, we will derive attenuation curves for the member galaxies as these are relevant to accurately estimate stellar masses and SFRs, and provide information on the dust properties \citep[e.g.,][]{shivaei20a,shivaei20b,polletta24}.\\
\textbf{(4) Kinematic analysis: }
VESPER spectral resolution (R$\sim$ 3000) will enable the identification of sub-components characterized by different velocity shifts, widths, and intensities. The rotation velocity and the velocity dispersion will be estimated to characterize the disk dynamics, search for non rotational components such as inflows or outflows, and look for signs of interactions, disk instabilities, gas accretion and ejection processes \citep[see e.g.,][]{morishita23,arribas24}.  The different sub-components will be separated and for each of them we will quantify the spatial extent, and the luminosity. A possible approach is to analyze the residuals from the surface brightness maps after subtracting axially symmetric components, such as a spheroid and a disk.  In particular, we will look for radially inflowing or outflowing gas \citep[see e.g.,][]{genzel23,forster_schreiber24} and measure its mass, and rate. The kinematics of low metallicity gas will be mapped as it might provide evidence of inflow of pristine gas.

\begin{figure*}
    \centering
    \includegraphics[width=0.45\textwidth]{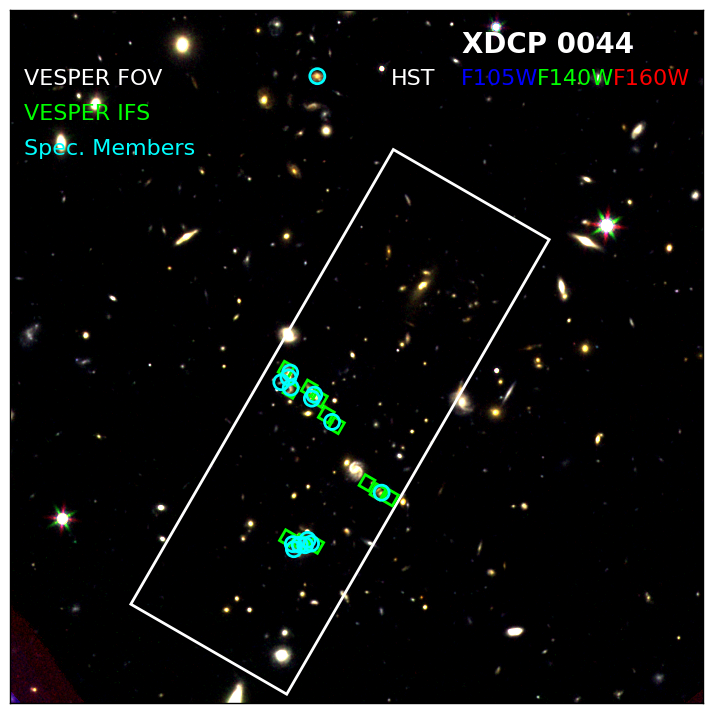}
    \includegraphics[width=0.45\textwidth]{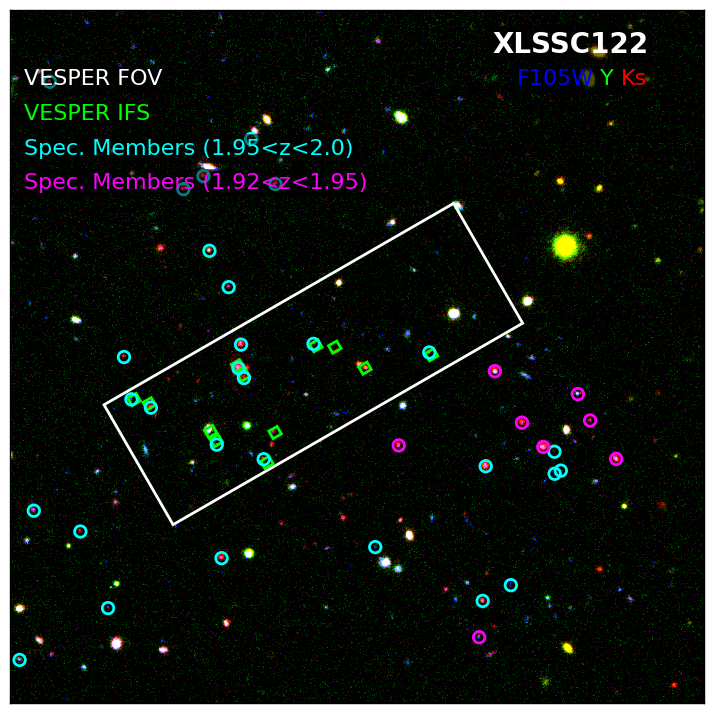}
   \caption{Images of the clusters XDCP\,J0044.3$-$2033 at
$z=1.58$ (left panel), and XLSSC\,122 at $z=1.99$ (right panel) with a possible positioning of the VESPER IFS (green rectangles). The full VESPER FOV (24\arcsec$\times$70\arcsec) is shown as a white rectangle. Spectroscopic members are indicated with cyan and magenta circles. }
    \label{fig:clusters}
\end{figure*}
\begin{figure*}%[ht]
    \centering
    \includegraphics[width=0.45\textwidth]{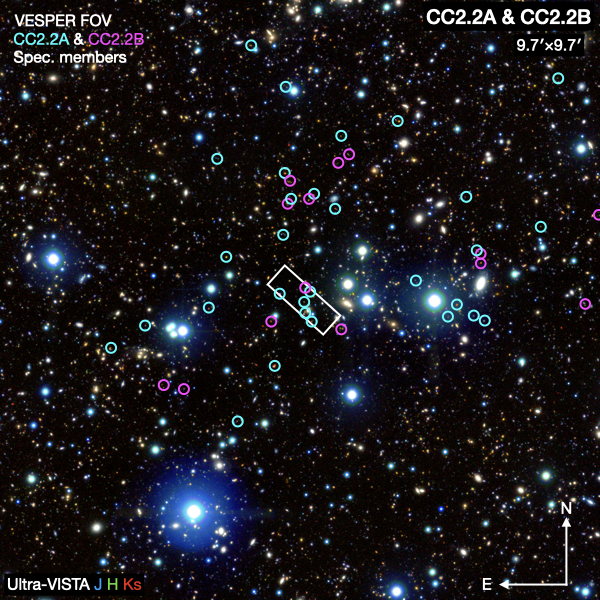}
    \includegraphics[width=0.45\textwidth]{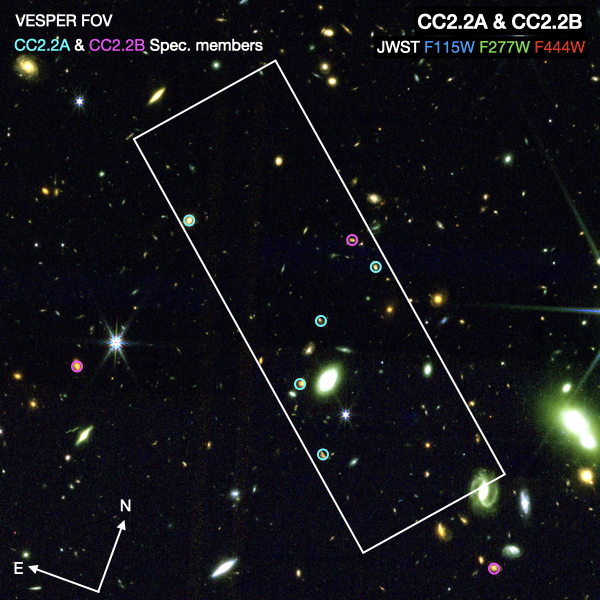}
   \caption{Near-infrared images of the galaxy protoclusters CC2.2A and CC2.2B \citep{darvish20,darvish24}. The left panel shows the Ultra-VISTA (JHKs) 9.7\arcmin$\times$9.7\arcmin\ image, and the right panel the 1.5\arcmin$\times$1.5\arcmin\ JWST (F115W, F277W, and F444W) image of the CC2.2A protocluster core. The spectroscopic members are shown as open circles (cyan for CC2.2A and magenta for CC2.2B).}
    \label{fig:sel_pc}
\end{figure*}

\subsection{Selected galaxy clusters and protoclusters}

Out targets are high redshift clusters and protoclusters at $1.5<z<2.5$. In this redshift range, VESPER wavelength coverage (1.2-2.4$\mu$m) will give access to the rest-frame visible spectrum (i.e., from 4800\AA\ to 7000\AA) and to major visible emission lines (i.e., \hbeta, \oiiib, \Ha, and \sii). To avoid having major spectral features in windows of low atmospheric transmission, we exclude sources at $1.7<z<2.0$. A list of potential candidates is reported in Table~\ref{tab:targets}, this includes six clusters at $1.5<z<1.7$ and eight protoclusters at $2.0<z<2.5$. They are all visible from the ELT site (i.e., $-70$\deg$<\delta<20$\deg). We will target the massive members of the densest region of each structure. Our choice will be refined based on the number of confirmed members, and on additional criteria such as the estimated halo mass and the availability of ancillary multi-wavelength data.  We do not rule out including systems that will be discovered in the near future, but we point out that there is already a good number of potential targets for this program. \\
The choice of observing five clusters and five protoclusters is justified by the need of capturing the range of member properties observed in different structures. Indeed they can exhibit quite different behaviors even if they are at the same redshift (see e.g., the different quiescent fractions in the clusters JKCS\,041 and CL\,J1449; \citealt{newman14,strazzullo18}, or the different starburst fractions in the protoclusters Spiderweb and USS\,1558-003; \citealt{perez_martinez23, perez_martinez24}). Furthermore, such a sample size well matches the number of known structures in each of the two pre-defined groups (clusters at $1.5<z<1.7$, and protoclusters at $2<z<2.5$; see Table~\ref{tab:targets}). Finally, with this sample size, the proposed program can be carried out with a reasonable request of observing time. 

\begin{table}[ht!]
\caption{Potential SHARP targets}\label{tab:targets}
\begin{tabular*}{\tblwidth}{@{}LCCCCC@{}}
\toprule
     Target name$^a$               & $\alpha$  & $\delta$ & $z$ &  N$^b$ & Ref.$^c$\\
                               & (h:m:s)   & (\deg:\arcmin:\arcsec) & &   &     \\
\midrule
Cluster & \multicolumn{5}{c}{$1.5<z<1.7$}\\
\midrule
 XDCP\,J0044                  & 00:44:05.2  & $-$20:33:58.0  & 1.58 &  22 & (1,2) \\
 ClG\,J0218                   & 02:18:21.0  & $-$05:10:30.0  & 1.62 &  21 & (3,4)  \\
 J1030+0524                   & 10:30:25.2  &  \005:24:28.7  & 1.69 &  10 & (5)\\ 
 JKCS\,041                    & 02:26:44.6  & $-$04:41:38.4  & 1.80 &  19 & (6,7) \\
 XLSSC\,122                   & 02:17:44.1  & $-$03:45:36.0  & 1.99 &  37 & (8)\\
 CL\,J1449                    & 14:49:14.0  &  \008:56:21.0  & 1.99 &  14 & (9,10) \\
\midrule
Protocluster & \multicolumn{5}{c}{$2.0<z<2.5$}\\
\midrule
 MRC\,0156$-$252              & 29:38:21.9  & $-$24:59:32.1  & 2.02 &  10 & (11) \\
 ZFIRE                        & 10:00:22.5  &  \002:15:56.3  & 2.09 &  41 & (12,13)  \\
 MRC\,1138-262                & 11:40:48.3  & $-$26:29:09.0  & 2.16 &$>$60 & (14--18) \\
 PHz\,G237                    & 10:01:51.6  &  \002:18:36.0  & 2.16 &  31 & (19,20) \\
 CC2.2A                       & 10:00:47.4  &  \002:00:11.6  & 2.23 &  35 & (21) \\
 CC2.2B                       & 10:01:26.0  &  \001:54:31.8  & 2.24 &  17 & (22)\\
 PCL1002                      & 10:00:25.1  &  \002:25:00.3  & 2.47 &  27 & (13,23)\\
 4C\,23.56                    & 21:07:14.8  &  \023:31:45.1  & 2.49 &  21 & (24--27) \\
\bottomrule
\end{tabular*}
\begin{flushleft}{$^a$ XDCP\,J0044: XDCP\,J0044$-$2033; ClG\,J0218: ClG\,J0218$-$ 0510; CL\,J1449: CL\,J1449$+$0856; PHz\,G237: PHz\,G237.01$+$ 42.5.
$^b$ N: number of spectroscopically confirmed members.
$^c$: References: 1: \citet{travascio20}, 2: \citet{lepore22}, 3: \citet{papovich10}, 4: \citet{rudnick17}, 5: \citet{damato20}, 6: \citet{andreon14}, 7: \citet{newman14}, 8: \citet{willis20}, 9: \citet{gobat11}, 10: \citet{coogan18}, 11: \citet{galametz13}, 12: \citet{hung16}, 13: \citet{zavala19}, 14: \citet{perez_martinez23}. 15: \citet{zhang24_spiderweb}, 16: \citet{shimakawa24}, 17: \citet{shimakawa25}, 18: \citet{naufal24}, 19: \citet{koyama21}, 20: \citet{polletta21}, 21: \citet{darvish20}, 22: \citet{darvish24}, 23: \citet{casey15}, 24:  \citet{tadaki19}, 25: \citet{zhou24}, 26: \citet{lee17}, 27: \citet{lee19}.}
\end{flushleft}
\end{table}

\subsection{Observing strategy}
\label{sec:strategy}

Our strategy is to observe the massive (M$_{\rm star}>10^{11}$\,\msun) galaxies in the densest regions of clusters and protoclusters at $1.5<z<2.5$. To carry out this program we need both a sufficiently wide field of view (FOV) to fully cover the core, enough spatial and spectral resolution to identify spatially and kinematically different sub-components within each galaxy member, and a wavelength coverage that  gives access to the main visible spectral features.\\
Based on current studies, high-$z$ clusters and protoclusters contain usually two, and in a few cases up to seven members with stellar mass $>$10$^{10.5}$\,\msun\ in a 10\arcsec\ region (see e.g., the cluster XDCP\,J0044.3$-$2033 at $z=1.58$, \citealt{travascio20}; the protocluster Distant Red Core (DRC) at $z=4$, \citealt{oteo18}; SPT J2349$-$5638 at $z=4.3$, \citealt{miller18}; and CLJ1001 at $z=2.51$, \citealt{xu25}). The core regions of cluster and protoclusters at the selected epoch  are typically $\lesssim$1\arcmin. VESPER FOV (24\arcsec$\times$70\arcsec) is thus well suited to the size of their cores. We anticipate to place about six IFS on confirmed members for each targeted core in the VESPER FOV. The remaining IFS will be placed on less massive member candidates or dedicated to investigate companions or extended features around the selected targets, such as tidal tails and bridges. Possible positionings of the VESPER FOV and of the 12 IFS in two clusters, and a protocluster are shown in Fig.~\ref{fig:clusters}, and~\ref{fig:sel_pc}, respectively. In all cases, from six to nine member galaxies fall within the VESPER FOV, effectively using the multiplex capabilities of the instrument. By assuming an average of six members per structure, we expect to have two samples of 30 targets (6 members in 5 structures), and 60 targets in total combining the cluster and protocluster subsamples. \\
In addition, VESPER offers the wavelength coverage ($1.2-2.4\,\mu$m) to observe the rest-frame visible spectrum and reveal moderately obscured regions for objects at $z\sim2$. Its spatial (1 pixel = 31\,mas) and spectral resolution ($R=3000$) are well suited to map the different galactic components and their kinematics in galaxies at $z\sim2$. 
Finally, its multiplexing capabilities maximize the observing efficiency when targeting dense fields.\\
Currently there are no facilities that can carry out the proposed program as effectively as VESPER. Other sensitive spectrographs, such as MOSAIC@ELT, do not have a sufficiently long wavelength coverage to enable the study of \Ha\ at $z>1.7$. None has the capability and FOV to observe simultaneously multiple targets, and is thus suitable for observing efficiently crowded regions as required for this study.  For example, the JWST NIRSpec integral field unit (IFU) provides spatially resolved imaging spectroscopy over a 3\arcsec$\times$3\arcsec\ region with exquisite angular (0.1\arcsec) and spectral (R$\sim1000-2700$) resolution over a broad wavelength range ($0.6-5.3\mu$m), but does not offer multiplexing capabilities.

\subsection{Depth and exposure time}
\label{sec:etc}
To estimate the depth necessary to carry out the analysis described above we consider the expected brightness of the stellar continuum, and of the \Ha\ emission line. For the continuum, we consider a limiting K-band magnitude of 22 which is typical of the targeted galaxies and corresponds to SFGs and quiescent galaxies at $1.5<z<2.5$ with stellar masses $>10^{11}$\,\msun\ \citep{weaver23}. 

Using the SHARP Exposure Time Calculator (ETC v0.6\footnote{https://sharp.lambrate.inaf.it\/}) and assuming a Sersic profile with index $n=1$ and effective radius 350\,mas \citep[$\sim$3\,kpc at $z=2$;][]{polletta24,vanderwel25}, we can reach a S/N at the reference wavelength (2.2$\mu$m) of $\sim$5 in the center
with an exposure of 8\,hr and by binning 8 spectral pixels (two resolution elements) and 3$\times$3 spatially. \\
The mass limit of $10^{11}$\,\msun\ corresponds to a SFR of 84$\pm$27 \msun\ yr$^{-1}$ assuming the star-forming main sequence (MS) relation at $z=2.0\pm0.5$ \citep{popesso23}. For a galaxy with this SFR, the expected \Ha\ luminosity, from star formation only and without taking extinction into account, would be $\sim1.6\pm0.5\times 10^{43}$\ergs, assuming the SFR--L(\Ha) relation in \citet{kennicutt12} and a Chabrier initial mass function \citep[IMF;][]{chabrier03}. At the sources redshift these luminosities correspond to total \Ha\ fluxes of $(4-7)\times10^{-16}$\ergcm2s. With 8\,hr exposure the ETC tells us that we would detect a line with flux of $4\times10^{-16}$\ergcm2s, and width of FWHM$\sim$1000\,\kms\ ($\Delta\lambda=65$\AA) with S/N$\sim$70 in the center and with S/N$\sim$30 at the effective radius at $\lambda=1.97\mu$m (corresponding to \Ha\ at $z=2$), assuming the same Sersic profile as for the continuum. 
These S/N are sufficient to carry out the spectral and kinematic analysis described in Sect.~\ref{sec:strategy}.\\
We thus request an average of 8\,hr of integration time per structure, but this will be adjusted according to the redshift in order to have a uniform mass limit across the sample.\\
In conclusion, with an average integration time of 8\,hr per source,
VESPER will deliver deep (S/N$>5$ in the continuum, and $\sim$70 in the \Ha\ line) spatially resolved maps with 0.1\arcsec\ angular resolution of the stellar and gaseous components of $\sim$60 massive members ($>10^{11}$\,\msun), and maps at slightly larger angular resolution of less massive members and extended structures such as bridges, tidal tails, and nearby objects situated in the cores of clusters and protoclusters across $z\sim2$. Thus, for a total of ten fields (5 clusters at $1.5<z<1.7$ and 5 protoclusters at $2.0<z<2.5$) and at least 60 members, we request 80\,hr$+$overheads of total integration time to carry out the proposed project. 

\section{Synergy with other facilities} 
\label{sec:synergy}
 
The proposed study will greatly benefit from current and forthcoming facilities expected to become available over the next decade. A comprehensive understanding of the evolution of structure members requires key observables probing cold and warm gas phases, AGN activity, and both local and global environments.

The cold molecular gas component can be mapped through high–angular-resolution observations of CO emission lines with the Atacama Large Millimeter Array (ALMA). These data will constrain the gas mass and origin, gas fractions, and star formation efficiencies, as well as the dynamical state of the systems. CO kinematics will also reveal signatures of inflows and outflows, providing direct insight into the physical mechanisms regulating star formation.

Warm gas and hot dust emission can be traced by JWST and the Roman Space Telescope (operational between 2027 and 2032), enabling the identification of heavily obscured and shock-heated regions.

AGN activity, which appears to be enhanced in overdense environments in both X-rays \citep{polletta21,vito24} and radio wavelengths \citep{chapman25}, can be characterized through MIR, radio, and X-ray observations. These diagnostics reveal circumnuclear hot dust, AGN-powered radio emission, and accretion-driven X-ray emission. Such studies will be enabled by future facilities including NewAthena (launch expected in 2037) and the Square Kilometre Array (SKA; full science operations expected by 2028). These observatories will also allow detailed investigations of the intracluster medium (ICM), which plays a fundamental role in processes such as gas accretion, strangulation, and ram-pressure stripping. In addition, SKA will map the distribution of atomic hydrogen around cluster members, providing critical constraints on large-scale environmental mechanisms.

Constraining the spatial extent and halo mass of the targeted structures requires sensitive wide-area spectroscopic coverage. Major contributions in this regard might come from the Vera C. Rubin Observatory and its Legacy Survey of Space and Time (LSST; expected to conclude by 2035), the Multi-Object Optical and Near-infrared Spectrograph (MOONS) at the VLT (operations beginning in 2026), and Euclid (operational through 2029). Together, these facilities will enable robust member identification and detailed characterization.

In summary, each facility provides unique and complementary diagnostic capabilities. Combined, they will deliver an unprecedented, multi-wavelength view of structures at $z\sim2$ and their galaxy populations, ultimately revealing their critical role as sites of massive galaxy formation.
\section*{Acknowledgments}

The SHARP team acknowledges support by Bando Ricerca Fondamentale INAF
2022, Techno-Grant ``SHARP’’ - 1.05.12.02.01 and Bando Ricerca Fondamentale
INAF 2024, Large-Grant ``SHARP’’ - 1.05.24.01.01.  MP acknowledges support
by Bando Ricerca Fondamentale INAF 2023, Mini-Grant - 1.05.23.04.01.

\printcredits

\bibliographystyle{cas-model2-names}

\end{document}